\documentclass[prl,twocolumn,a4paper]{revtex4}
\usepackage{graphicx,amsmath}
\usepackage{subfigure}
\usepackage{times}

\newcommand\op[1]{\hat{#1}}
\newcommand\vek[1]{\text{\textit{\textbf{#1}}}}
\newcommand{\bra}[1]{\mathord{\left \langle#1 \right|}}
\newcommand{\ket}[1]{\mathord{\left| #1 \right\rangle}}
\newcommand\Ca{$^{40}$Ca$^+$}
\newcommand\schema[3][0.2\textwidth]{%
	\subfigure[][
		\begin{minipage}{#1}\scriptsize\baselineskip1.5em
			#3
		\end{minipage}
		]{\includegraphics[width=0.175\textwidth]{#2}}%
}
\begin{document}

\title{Single calcium-40 ion as quantum memory for photon polarization: a case study}
\date{\today}
\author{Philipp M\"uller}
\author{J\"urgen Eschner}
\affiliation
{Experimentalphysik, Universit\"at des Saarlandes, Campus E\,2\,{\scriptsize6}, 66123 Saarbr\"ucken, Germany}

\begin{abstract}
\noindent We present several schemes for heralded storage of polarization states of single photons in single ions, using the \Ca\ ion and photons at 854\,nm wavelength as specific example. We compare the efficiencies of the schemes and the requirements for their implementation with respect to the preparation of the initial state of the ion, the absorption process and its analysis. These schemes may be used to create and herald entanglement of two distant ions through entanglement swapping; they are easily adapted to other atomic systems and wavelengths.
\end{abstract}

\maketitle

\noindent The controlled transfer of quantum states between different systems is a cornerstone in the realization of quantum networks. One proposed implementation~\cite{cirac1997} of such a network employs single atoms for storing information (\emph{i.\,e.}, as nodes of the network) and single photons for transmitting information (\emph{i.\,e.}, as channels of the network); they are interfaced through controlled absorption and emission. 

In atom--photon interaction, the conservation of angular momentum relates the internal state of the atom with the polarization state of the exchanged photon. In emission, this generates atom--photon entanglement which may be used to prepare a certain superposition of Zeeman sublevels in the atom by measuring an emitted photon in a certain polarization basis \cite{blinov2004, rosenfeld2007}, or to preserve such a superposition state in the context of quantum error correction \cite{akerman2012}. In absorption, a certain polarization state of the photon excites the absorbing atom into the corresponding superposition of Zeeman sublevels. Using neutral rubidium\nobreak-87 in this way as a quantum memory for an arbitrary polarization state of a photon at 780\,nm has been proposed by \cite{lloyd2001} and put into effect by \cite{specht2011, ritter2012}, using optical cavities to attain high efficiency.

Here, we propose schemes employing both absorption and emission to implement \emph{heralded} storage of the polarization state of a photon. First, the atom is prepared in a pure state ready to absorb a single photon. The absorption of a (certain but unknown) polarization state excites the atom into a corresponding superposition. Afterward, the excited atomic levels decay to different final levels, and another photon is emitted, which is entangled with the atom. By choosing the proper detection basis for this photon, the emission channels are made indistinguishable and the detection heralds the storage process but does not reveal the stored information. Due to the heralding, high fidelity will be achieved independent of the efficiency of the process.
We describe in detail the requirements and conditions for such heralded quantum storage processes for the case of singly ionized calcium\nobreak-40, the species used in our~\cite{schug2013, kurz2013, huwer2013, piro2011} and other experiments~\cite{FSK, drewsen, keller, knoop, blatt}. Application to other species is straightforward. 

\begin{figure}[htb]
\includegraphics[width=\linewidth]{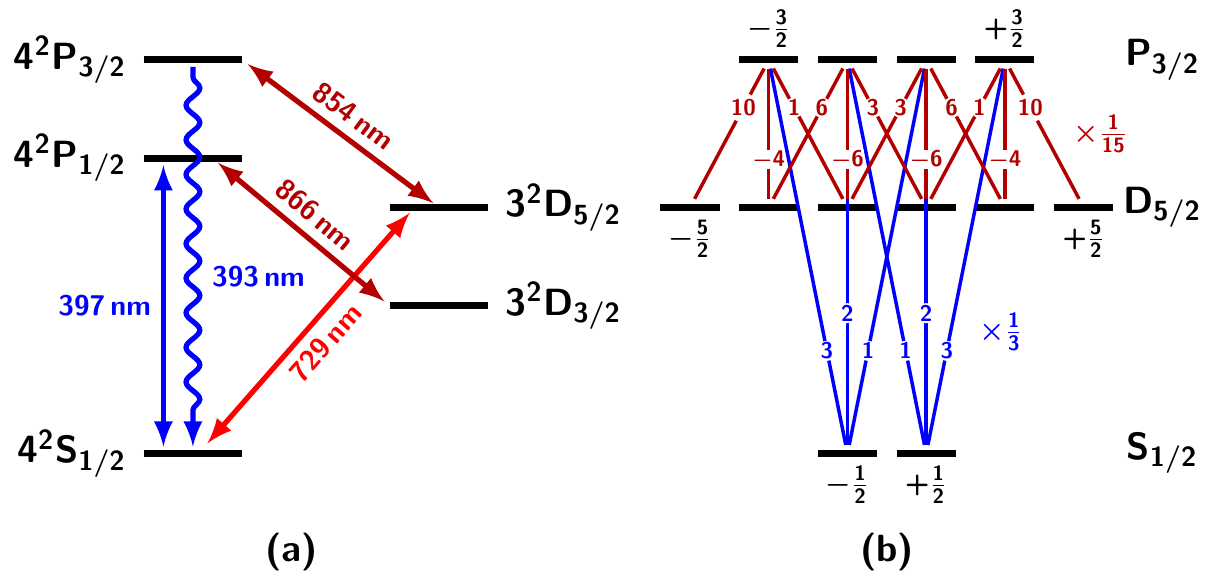}
\caption{(a) Level scheme and transitions for the \Ca\ ion. The branching ratios for the decay of P$_{3/2}$ are 94\,\% into S$_{1/2}$, 6\,\% into D$_{5/2}$ and $<$\,1\,\% into D$_{3/2}$. (b) Clebsch--Gordan coefficients (CGC); the CGC of a particular $j, m \rightarrow j', m'$ transition is obtained by multiplying the modulus of the respective number with the factor on the right, taking the square root and applying the sign indicated with the number.\label{levels}}
\end{figure}

In the proposed quantum-memory schemes, a single \Ca\ ion is used to store the polarization state of a single photon at 854\,nm wavelength (\emph{cf.}\ fig.\,\ref{levels}\,a): the absorption happens on the 3\,D$_{5/2}$--4\,P$_{3/2}$ transition and the emission of the herald takes place on the 4\,P$_{3/2}$--4\,S$_{1/2}$ transition at 393\,nm. Thereby the photonic qubit state is mapped to the atomic qubit formed by the Zeeman sublevels of the S$_{1/2}$ ground state.

\begin{figure}[htb]
\includegraphics[width=0.85\linewidth]{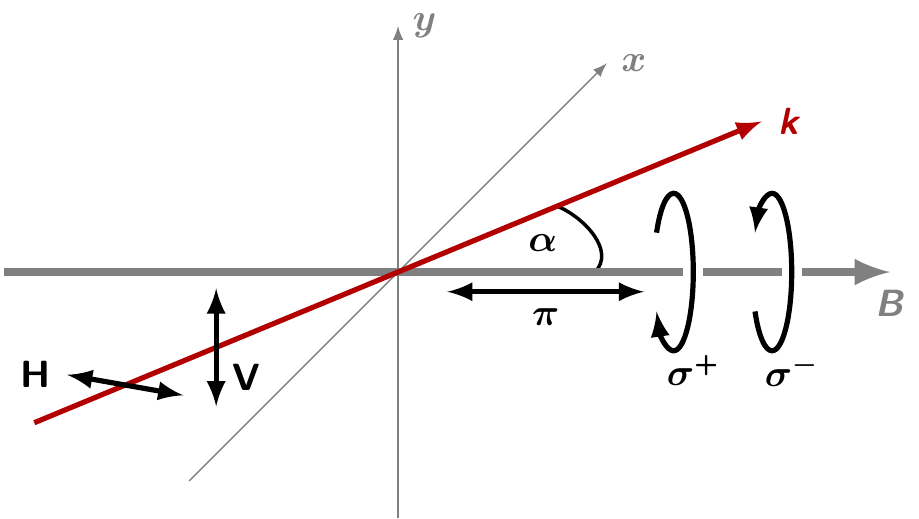}
\caption{Geometry of the polarizations (see text). 
\label{geometrie}}
\end{figure}

In general, the procedure begins with preparing the ion in a pure state $\ket{\psi_\text D} = \sum c_{m_\text D} \ket{m_\text{D}}$ in the D$_{5/2}$ multiplet. The photonic qubit is given by the polarization state of the 854-nm photon, which we express as a superposition of two orthogonal linear polarizations, say H and V,
\begin{equation*}
\ket{\phi_{854}} = \cos\vartheta\ket{\text H} + \text e^{\text i\varphi}\sin\vartheta\ket{\text V}.
\end{equation*}
How these polarizations translate into the individual atomic transitions (\emph{cf.}\ fig.\,\ref{levels}\,b) depends on the angle $\alpha$ between the propagation direction $\vek k$ of the 854-nm beam and the quantization axis, defined by an external magnetic field $\vek B$ (see fig.\,\ref{geometrie}). 
With H being the polarization in the plane of $\vek k$ and $\vek B$, this is done by applying the unitary transformation
\begin{equation*}
\textstyle\op P_\alpha = \left( \sin\alpha \ket0 {+} \cos\alpha\ket x \right)\bra{\text H} + \ket y\bra{\text V},
\end{equation*}
where $\ket0$ and $\ket{\pm 1} = (\ket x \pm \text i\ket y)/\sqrt2$ are the natural polarizations of the atomic transitions, corresponding to $\Delta m = 0$~($\pi$) and $\Delta m = \pm 1$~($\sigma^\pm$) transitions, respectively.

The absorption process is governed by the operator
\begin{equation*}
\op A = \sum_{\hbox to 0pt{\hss$\mathsurround=0pt{\scriptstyle m_\text D, m_\text P}$\hss}} C_{m_\text D, m_{854}, m_\text P} \ket{m_\text P} \bra{m_\text D} \bra{m_{854}},
\end{equation*}
where $C_{m_\text D, m_{854}, m_\text P}$ are the Clebsch--Gordan coefficients \cite{CG} displayed in fig.\,\ref{levels}\,b and $m_{854} = m_\text P - m_\text D = 0, \pm1$ determines the polarizations of the individual transitions.

In a similar way, the emission operator is given by
\begin{equation*}
\op E = \sum_{\hbox to 0pt{\hss$\mathsurround=0pt{\scriptstyle m_\text S, m_\text P}$\hss}} C_{m_\text P, m_{393}, m_\text S} \ket{m_\text S} \ket{m_{393}} \bra{m_\text P}
\end{equation*}
whereby $m_{393} = m_\text P - m_\text S = 0, \pm1$ determines the polarizations of the emission.

This procedure transforms the initial state into a joint (and in general entangled) state of the ion in the S$_{1/2}$ manifold and the emitted 393-nm photon,
\begin{equation*}
\ket{\psi_\text{joint}} = \op E \op A (\ket{\psi_\text D} \otimes \op P_\alpha \ket{\phi_{854}}).
\end{equation*}
After detecting the 393-nm photon at the angle $\alpha'$ to the quantization axis and with a polarization
\begin{equation*}
\ket{\phi_\text{det}} = \cos\vartheta'\ket{\text H} + \text e^{\text i\varphi'}\sin\vartheta'\ket{\text V},
\end{equation*}
the ion ends up in the state $\ket{\psi_\text S} = \bra{\phi_\text{det}} \op P_{\alpha'}^\dagger \ket{\psi_\text{joint}}$, which depends on $\vartheta$, $\varphi$, $\alpha$, $\alpha'$, $\vartheta'$, $\varphi'$ and on the initial atomic state $\ket{\psi_\text D}$. With this formula, one determines the mapping $\ket{\phi_{854}} \to \ket{\psi_\text S}$ and the relative efficiency~$\varepsilon$ of the storage process, which we define as the square of the modulus of $\ket{\psi_\text S}$ divided by the largest possible value, 2/3 in this case. To keep the full information of the qubit, it has to be ensured that $\varepsilon$ is independent of the initial photonic state and that orthogonal initial states (like $\ket{\text H}$ and $\ket{\text V}$) are mapped to orthogonal final states in the ion. To hold these constraints, one has to properly choose the directions of absorption and detection ($\alpha$ and $\alpha'$) as well as the detection basis ($\vartheta'$ and $\varphi'$).

\begin{widetext}\centering
\begin{figure}[htb]
\begin{minipage}{\textwidth}
\schema{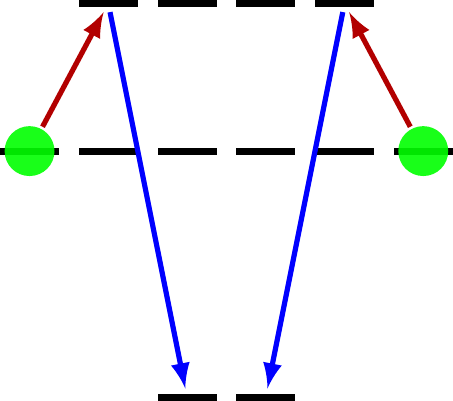}{%
	$\ket{\psi_\text D} = \frac{1}{\sqrt2}(\ket{-\frac{5}{2}} + \ket{+\frac{5}{2}})$\\
	$\alpha = 0$, 
	$\alpha' = 0$\\
	$\vartheta' = \varphi' = 0$ (H and V)\\
	$\ket{\text H} \to \frac{1}{\sqrt2}(\ket{+\frac{1}{2}} \pm \ket{-\frac{1}{2}})$\\
	$\ket{\text V} \to \frac{1}{\sqrt2}(\ket{+\frac{1}{2}} \mp \ket{-\frac{1}{2}})$\\
	$\ket{\text R} \to \ket{-\frac{1}{2}}$\\
	$\ket{\text L} \to \ket{+\frac{1}{2}}$\\
	$\varepsilon = 50\,\%$}
\hfill
\schema{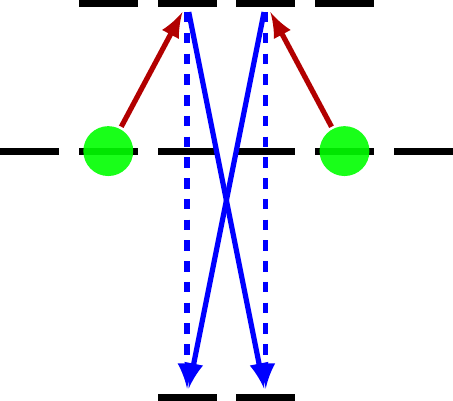}{%
	$\ket{\psi_\text D} = \frac{1}{\sqrt2}(\ket{-\frac{3}{2}} + \ket{+\frac{3}{2}})$\\
	$\alpha = 0$, 
	$\alpha' = 90^\circ$\\
	$\vartheta' = \varphi' = 0$ (H and V)\\
	$\ket{\text H} \to \frac{1}{\sqrt2}(\ket{+\frac{1}{2}} \pm \ket{-\frac{1}{2}})$\\
	$\ket{\text V} \to \frac{1}{\sqrt2}(\ket{+\frac{1}{2}} \mp \ket{-\frac{1}{2}})$\\
	$\ket{\text R} \to \ket{\mp\frac{1}{2}}$\\
	$\ket{\text L} \to \ket{\pm\frac{1}{2}}$\\
	$\varepsilon = 25\,\%$}
\hfill
\schema{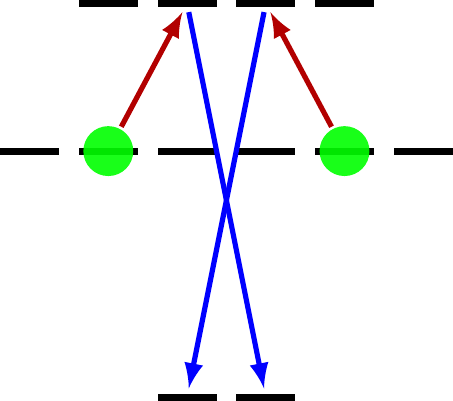}{%
	$\ket{\psi_\text D} = \frac{1}{\sqrt2}(\ket{-\frac{3}{2}} + \ket{+\frac{3}{2}})$\\
	$\alpha = 0$, 
	$\alpha' = 0$\\
	$\vartheta' = \varphi' = 0$ (H and V)\\
	$\ket{\text H} \to \frac{1}{\sqrt2}(\ket{+\frac{1}{2}} \pm \ket{-\frac{1}{2}})$\\
	$\ket{\text V} \to \frac{1}{\sqrt2}(\ket{+\frac{1}{2}} \mp \ket{-\frac{1}{2}})$\\
	$\ket{\text R} \to \ket{+\frac{1}{2}}$\\
	$\ket{\text L} \to \ket{-\frac{1}{2}}$\\
	$\varepsilon = 10\,\%$}
\hfill
\schema{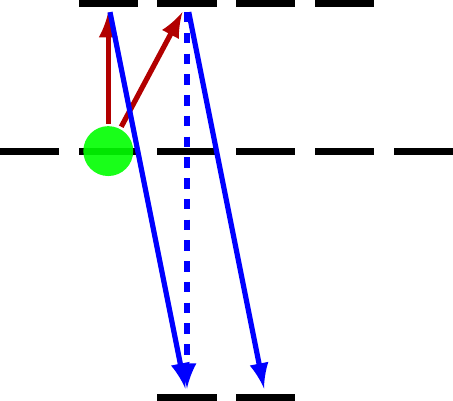}{%
	$\ket{\psi_\text D} = \ket{-\frac{3}{2}}$\\
	$\alpha = 47.06^\circ$,
	$\alpha' = 47.06^\circ$\\
	$\vartheta' = -55.74^\circ, \varphi' = \frac{\pi}{2}$\\
	$\ket{\text H} \to 0.56\ket{+\frac{1}{2}} - 0.83\ket{-\frac{1}{2}}$\\
	$\ket{\text V} \to 0.83\ket{+\frac{1}{2}} + 0.56\ket{-\frac{1}{2}}$\\
	$\ket{\text R} \to 0.98\ket{+\frac{1}{2}} - 0.19\ket{-\frac{1}{2}}$\\
	$\ket{\text L} \to 0.19\ket{+\frac{1}{2}} + 0.98\ket{-\frac{1}{2}}$\\
	$\varepsilon = 10.72\,\%$}
\hfill
\schema{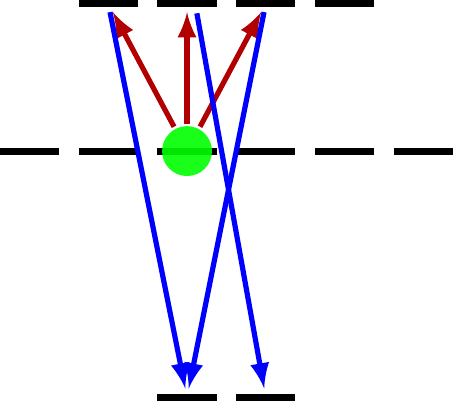}{%
	$\ket{\psi_\text D} = \ket{-\frac{1}{2}}$\\
	$\alpha = 90^\circ$, 
	$\alpha' = 90^\circ$\\
	$\vartheta' = 90^\circ$ (only V)\\
	$\ket{\text H} \to \ket{+\frac{1}{2}}$\\
	$\ket{\text V} \to \ket{-\frac{1}{2}}$\\
	$\ket{\text R} \to \frac{1}{\sqrt2}(\ket{+\frac{1}{2}} + \ket{-\frac{1}{2}})$\\
	$\ket{\text L} \to \frac{1}{\sqrt2}(\ket{+\frac{1}{2}} - \ket{-\frac{1}{2}})$\\
	$\varepsilon = 10\,\%$}
\caption{\label{schemes}Chart of possible schemes: \emph{pictures} show the level schemes (according to fig.\,\ref{levels}\,b) with prepared initial state and absorption and emission channels. Beneath are listed the initial atomic state, the angles of the geometry and of the detection basis, the mapping and the relative efficiency. See text for more details. The states $\ket{\text R} = \smash{\frac{1}{\sqrt 2}}(\ket{\text H} + \text i \ket{\text V})$ and $\ket{\text L} = \smash{\frac{1}{\sqrt 2}}(\ket{\text H} - \text i \ket{\text V})$ are the two circular polarization states. In (a)--(c) both emitted polarizations, H and V, serve as herald; the $\pm$ sign refers to the two possible detections.}
\end{minipage}
\end{figure}
\end{widetext}

Figure~\ref{schemes} illustrates some possible schemes with the prepared initial state $\ket{\psi_\text D}$ and the values of $\alpha$, $\alpha'$, $\vartheta'$ and $\varphi'$. For each case, the resulting mapping of the two orthogonal linear polarization states, $\ket{\text H}$ and $\ket{\text V}$, and of the two circular polarization states, $\ket{\text R}$ and $\ket{\text L}$, onto the ion's S$_{1/2}$ sublevels is displayed, as well as the relative storage efficiency. The highest efficiency is reached with scheme (a), where both in absorption and in emission, the selected channels are the only allowed transitions, and where both possible polarizations of the herald can be used. 
\\
\indent The decision which scheme is the most suitable to be implemented in an experiment depends not only on the efficiency~$\varepsilon$ but also on the experimental requirements. One issue is the preparation of the initial atomic state. To achieve a pure (single or superposition) state in D, the ion has to first be optically pumped into a single Zeeman sublevel, say to $\text S_{1/2}, m = -\frac 1 2$. Afterward, this state is transferred to the D$_{5/2}$ manifold.  In the simplest case of schemes (d) and (e), this is done by a single $\pi$-pulse of a narrow-band 729-nm laser driving a single branch of the S$_{1/2}$--D$_{5/2}$ quadrupole transition (note that the individual S--D transitions are frequency-resolved due to the applied magnetic field). For the other schemes, a $\pi/2$-pulse and a subsequent $\pi$-pulse on different branches have to be applied. These must be phase-locked pulses to sustain coherence of the superposition. For scheme~(a), an additional pulse at radio frequency is needed to flip the spin in the S$_{1/2}$ manifold because the selection rules for electric quadrupole transitions allow only $|\Delta m| \leq 2$. Since every pulse consumes time and may cause impurities in the system, all this unfolds a trade-off between the complexity of the state preparation and the relative efficiency of the storage process.

Another issue is the direction of the $\vek k$-vectors of the incoming and outgoing photons with respect to the magnetic field, given by the angles $\alpha$ and $\alpha'$. In our set-up \cite{schug2013, kurz2013, huwer2013, piro2011}, we use high-aperture laser objectives (HALO) for optimizing the single-photon absorption and detection efficiency. As both HALOs are on one axis, the condition $\alpha = \alpha'$ is preferred, which is satisfied by the schemes (a) and (c)--(e). Scheme (d) poses another practical challenge through the non-trivial angle $\alpha = \alpha' \approx 47^{\circ}$ between photonic $\vek k$-vector and magnetic field.

It is important to note that the arrival time of the heralding photon plays a role in the schemes, as a consequence of the energy splitting of the Zeeman levels which leads to Larmor precession of the involved coherent superpositions, \emph{i.\,e.}, to rotating phases. As has been shown in \cite{Lukin}, precise measurement of the detection time of the herald, with resolution much better than the Larmor precession time, is required to be able to determine these phases in the final atomic qubit state. 

The heralding through an emitted single photon enables high fidelity (\emph{i.\,e.}, in principle unity) of this quantum memory, not limited by the efficiency of absorption and detection. But uncertainties in the angles of direction and polarization of the to-be-stored photon and the heralding detection basis, as well as impurities of the initially prepared atomic state, may reduce the fidelity of the state mapping. We elaborate on some of these issues in the following. Specific numbers for the error sensitivity of the individual schemes may be calculated from their numerical modeling based on the methods explained above.

With respect to angular deviations, the most efficient scheme, fig.\,\ref{schemes}\,(a), is particularly robust because there are no competing transitions with unwanted polarization, neither in absorption nor in emission. There is a residual sensitivity to polarization mixing, however, through the finite solid angle used for collecting the heralding photons. In other schemes such as (b), (c) and (e), impurities of the incoming or of the detected polarization may admix unwanted transitions and thereby reduce the mapping fidelity  proportionally. Scheme~(d) is sensitive to angular deviations because it involves the destructive interference of scattering channels. In general, less efficient schemes, where unwanted transitions have to be suppressed through directional selection or polarization filtering, will be more error-prone.

In schemes where the initially prepared atomic state is a superposition, fig.\,\ref{schemes}\,(a)--(c), impurity of this state enters directly into the fidelity of the mapping. Such impurity may be caused by phase noise on the preparing laser or by fluctuating magnetic fields. State-of-the-art experimental technology allows for creating superposition states with very small impurity \cite{impurities}, as well as with millisecond coherence times \cite{monz2011}.

The presented schemes are the same for single-electron atoms with a metastable D level and vanishing nuclear spin, such as the even isotopes of Ca$^+$, Sr$^+$, Ba$^+$, Ra$^+$ and Hg$^+$. Moreover, the general method is easily applied to further atomic systems, including cases where absorption and emission are on the same transition, as well as atoms or atomic ions with non-zero nuclear spin. The latter offer using ground-state hyperfine levels as initial and final states, and they allow devising symmetric schemes with a single initial state ($m = 0$), as initially proposed in \cite{lloyd2001} and implemented in \cite{rosenfeld2007, specht2011}. Simultaneous application of one of the schemes in two distant atoms allows for their entanglement through the absorption of partner photons from an entangled photon pair~\cite{belltest}. 

\paragraph{Acknowledgements}
The authors wish to thank Nicolas Sangouard for motivating discussions. This work was partially supported by the BMBF (Verbundprojekt QuOReP, CHIST-ERA project QScale), the German Scholars Organization\slash Alfried Krupp von Bohlen und Halbach-Stiftung, the EU (AQUTE Integrating Project), and the ESF (IOTA COST Action).


\begin{thebibliography}{3}
\bibitem{CG} Here we use a suitably abbreviated notation for the CGCs. Various ``standard'' notations are found in the literature, to which out definition is related through $C_{m_\text D, m_{854}, m_\text P} = C_{m_\text D, m_{854}, m_\text P}^{5/2, 1, 3/2} = \left\langle 5/2, m_\text D; 1, m_{854} | 5/2, 1, 3/2, m_\text P \right\rangle$, and likewise for $C_{m_\text P, m_{393}, m_\text S}$.
\end{thebibliography}
\end{document}